\documentclass[twocolumn,preprintnumbers,amsmath,amssymb]{revtex4}

\usepackage{graphicx}
\usepackage{bm}

\begin{document}

\def\mstar{m_{\displaystyle *}}


\title{Autoresonant control of the many-electron dynamics in nonparabolic quantum wells}
\author{G. Manfredi and P.-A. Hervieux}
\affiliation{Institut de Physique et Chimie des Mat{\'e}riaux, CNRS
and Universit{\'e} Louis Pasteur, BP 43, F-67034 Strasbourg, France}

\date{\today}
\begin{abstract}
The optical response of nonparabolic quantum wells is dominated by
a strong peak at the plasmon frequency. When the electrons reach
the anharmonic regions, resonant absorption becomes inefficient.
This limitation is overcome by using a chirped laser pulse in the
autoresonant regime. By direct simulations using the Wigner
phase-space approach, we prove that, with a sequence of just a few
pulses, electrons can be efficiently detrapped from a nonparabolic
well. For an array of multiple quantum wells, we can create and
control an electronic current by suitably applying an autoresonant
laser pulse and a slowly varying dc electric field.
\end{abstract}

\maketitle

Small semiconductor devices, such as quantum dots and quantum
wells, have attracted considerable attention in recent years,
particularly for possible applications in the emerging field of
quantum computing \cite{Zoller}. For quantum devices working with
many electrons \cite{Muller}, it is crucial to understand the
properties of the self-consistent electron dynamics and its
response to external electric fields. Particular attention has
been devoted to intersubband transitions in semiconductor quantum
wells, which take place on the meV energy scale and involve
excitation frequencies of the order of the teraherz \cite{Heyman}.
Several theoretical and computational studies have investigated
the electron response, mainly using Hartree-Fock semiconductor
Bloch equations \cite{Nikonov} or density functional theory (DFT)
\cite{Ullrich}. For perfectly parabolic confinement, the electron
response is dominated by the Kohn mode \cite{Kohn, Dobson},
consisting of rigid oscillations of the electron gas at the
effective plasmon frequency. For nonparabolic confinement, the
Kohn mode still dominates the initial response. However, when the
electrons reach the anharmonic regions, the resonance condition is
lost and absorption becomes inefficient.

In this Letter, we show that this limitation can be overcome by
resorting to {\em autoresonant} excitation \cite{Fajans, Peinetti,
Marcus}. Basically, autoresonant excitation occurs when a
classical nonlinear oscillator is excited by an oscillating force
with slowly varying frequency:
$ F(t) = \epsilon \cos\left[\omega_0(t-t_0)+\frac{1}{2} \alpha
(t-t_0)^2\right]$,
where $\epsilon$ is the excitation amplitude and
$\omega_0$ the frequency, which matches the linearized oscillator
frequency; $\alpha$ is the rate of variation of the excitation
frequency.
For $|\alpha| \ll \omega_0^2$ and $\epsilon$ above a certain
threshold, the instantaneous oscillator frequency becomes "locked"
to the instantaneous excitation frequency, so that the resonance
condition is always satisfied. In that case, the amplitude of the
oscillations grows indefinitely and without saturation, until of
course some other effect kicks in. For the single-particle case
the threshold behaves as $\epsilon_{\rm th} \sim |\alpha|^{3/4}$
\cite{Fajans}.

In order to study the self-consistent electron dynamics, we make
use of the Wigner representation of quantum mechanics. A mixed
quantum state is represented by a function of the phase space
variables plus time: $f(x,v,t)$ (we deal with one-dimensional
problems), which evolves according to the Wigner equation
\begin{eqnarray}
&&\frac{\partial{f}}{\partial{t}} +
v\frac{\partial{f}}{\partial{x}}+\frac{i\mstar}{2\pi\hbar}
\int\int{d\lambda}~{dv'}e^{i\mstar(v-v')\lambda} f(x,v',t) \times \nonumber\\
&& \left[V\left(x+\frac{\lambda \hbar}{2},t\right)-
V\left(x-\frac{\lambda \hbar}{2},t\right)\right] =
\left(\frac{\partial{f}}{\partial{t}}\right)_{\rm scatt},
\label{Wigner}
\end{eqnarray}
where $\mstar$ is the effective electron mass and $V(x,t)$ is the
total potential acting on the electrons. The latter is composed of
three terms: (i) the confining potential $V_{\rm conf}(x)$; (ii)
the autoresonant oscillating potential $V_{\rm auto} = x F(t)$;
and (iii) the Hartree potential $V_H(x,t)$, which obeys Poisson's
equation
$V_H'' = (e^2/\varepsilon) \int_{-\infty}^\infty f\,dv $,
where $e$ is the absolute electron charge and $\varepsilon$ is the
effective dielectric constant. We consider wide quantum wells
($\approx 100$nm) at moderate electron temperatures ($\approx
2T_F$), for which the exchange and correlation corrections can be
neglected \cite{Santer, Gusev}.

The right-hand side of Eq. (\ref{Wigner}) models disorder or
phonon scattering in the form of a friction-diffusion term
\cite{Zurek}:
\begin{equation}
\label{scatt} \left(\frac{\partial{f}}{\partial{t}}\right)_{\rm
scatt} = 2\gamma \frac{\partial{(vf)}}{\partial{v}} + D_v
\frac{\partial^2{f}}{\partial{v^2}}+ D_x
\frac{\partial^2{f}}{\partial{x^2}}~,
\end{equation}
where $\gamma$ is the relaxation rate (inverse of the relaxation
time $T_1$), and $D_v$, $D_x$ are diffusion coefficients in
velocity and real space respectively, which are related to the
decoherence time $T_2$. In order for Eq. (\ref{Wigner}) to
preserve the positivity of the density matrix associated to the
Wigner distribution function, the scattering term must be in
Lindblad form \cite{Zurek}. This is automatically achieved
\cite{Isar} if the above coefficients respect the inequality $D_v
D_x \geq \gamma^2 \hbar^2/4\mstar^2$.

We focus on confining potentials that can be approximated by a
parabola at the bottom of the well:
$V_{\rm conf}(x) \simeq \frac{1}{2}\omega_0^2 \mstar x^2 + \dots$,
where the frequency $\omega_0$ can be related to a fictitious
homogeneous positive charge of density $n_0$ via the relation
$\omega_0^2 = e^2 n_0/\mstar \epsilon$. We then normalize time to
$\omega_0^{-1}$; space to the harmonic oscillator length $L_{\rm
ho}=\sqrt{\hbar/\mstar\omega_0}$; velocity to
$\sqrt{\hbar\omega_0/\mstar}$; energy to $\hbar\omega_0$; and the
electron density to $n_0$.

As initial condition, we take a Maxwell-Boltzmann velocity
distribution with temperature $T_e>T_F$ and a Gaussian density
profile with peak density $n_{e}$. This is not an exact stationary
state, but it evolves very little if no perturbation is applied.
(The precise form of the initial state is irrelevant for our
purposes, provided it is localized at the bottom of the well.)
We define the filling fraction as $\eta = n_{e}/n_0 \leq 1$: the
limit case $\eta=0$ corresponds to very dilute densities, for
which the Hartree potential is negligible.

We use typical parameters for semiconductor quantum wells
\cite{Ullrich}: effective electron mass and dielectric constant
$\mstar=0.067m_e$ and $\varepsilon = 13\varepsilon_0$; volume
density $n_0 = 5 \times 10^{16}{\rm cm}^{-3}$, $\omega_0 = 1.35
\times 10^{13}{\rm s}^{-1}$, $\hbar\omega_0 = 8.9 {\rm meV}$,
$L_{\rm ho}=11.3{\rm nm}$. For $\eta=1$, this yields a maximum
surface density for the electrons $n_s = 1.35 \times 10^{11}{\rm
cm}^{-2}$ and a maximum Fermi temperature $T_F = 85.7{\rm K}$. The
electron temperature is taken to be $T_e=2\hbar\omega_0 \simeq
200{\rm K}$. The relaxation time is $T_1 \approx 70$ps, which
corresponds to a relaxation rate $\gamma/\omega_0=0.001$. The
diffusion coefficient in velocity space is $D_v = \gamma \sqrt{k_B
T_e/\mstar}$.

First, we consider a single quantum well with a Gaussian
potential: $V_{\rm conf}(x) = -V_0\exp(-x^2/2\sigma^2)$, with
$V_0= \sigma^2\mstar\omega_0^2$, $\sigma=4L_{\rm ho}\simeq 45{\rm
nm}$, and an overall width of the quantum well of $L=24L_{\rm
ho}\simeq 270{\rm nm}$. This is in line with recent experiments on
wide parabolic quantum wells (WPQWs), which can reach a width of
several hundred nanometers \cite{Gusev}.

The electron gas is excited with a chirped pulse, with
$\alpha=-0.001 \omega_0^2$, $t_0 = 500 \omega_0^{-1}$, and
amplitude $\epsilon = 0.2 > \epsilon_{\rm th}$, in normalized
units. The latter corresponds to an electric field of the order of
0.1mV/nm, which can be easily achieved experimentally. Chirped
pulses were also suggested to excite transitions in two-level
quantum systems \cite{Batista}: a situation quite different from
the wide quantum wells considered here, where many levels are
present and progressively excited by means of the autoresonant
technique.

\begin{figure}[htb]
\includegraphics[height=5.5cm, width=7.5cm]{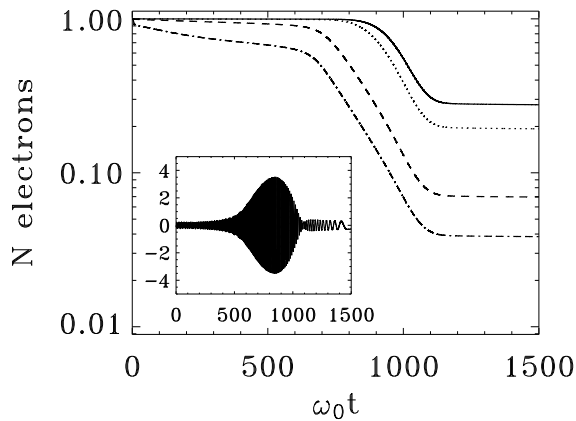}
\caption{\label{fig:1}Single Gaussian well. Total number of
electrons in the well, normalized to its value at $t=0$, for
different values of the filling fraction: $\eta=0$ (solid line),
$\eta=0.1$ (dotted), $\eta=0.5$ (dashed), $\eta=1$ (dot-dash).
Inset: Time evolution of the dipole for $\eta=0.5$.}
\end{figure}

Figure 1 shows the dipole of the electron distribution (average of
$x$) and the total number of electrons in the well (integral of
$f(x,v,t)$ in the phase space) normalized to its initial value,
for different values of the filling fraction $\eta$. The dipole
starts increasing rapidly at $t=t_0$, i.e. when the time-dependent
frequency of the laser field crosses the linear resonance
$\omega_0$. At $\omega_0 t \simeq 900$, the electrons that have
been accelerated by the laser field are finally ejected from the
well, leaving behind a remnant population that is still sitting at
the bottom of the well. The dipole then decreases rapidly and
remains close to zero thereafter. The number of electrons in the
well is reduced of one order of magnitude at the end of the run.
The slight initial decrease observed for large electron densities
($\eta=1$) represents a small evaporation due to Coulomb
repulsion.

\begin{figure}[htb]
\includegraphics[height=5cm, width=7.cm]{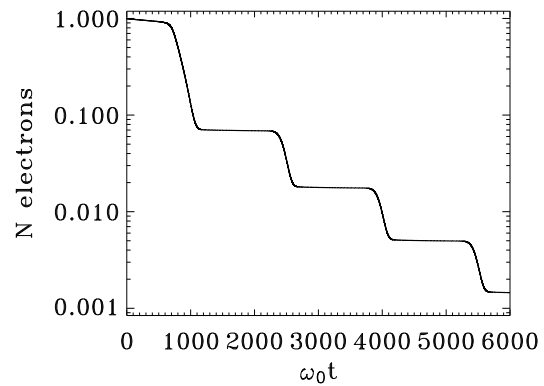}
\caption{\label{fig:2} Time evolution of the total number of
electrons for a series of four laser pulses; $\eta=0.5$.}
\end{figure}

By applying a sequence of four identical pulses it is possible to
reduce the number of electrons by almost {\em three orders of
magnitude}, as shown in Fig. 2. We stress that, for a non-chirped
pulse ($\alpha=0$) or for a chirped pulse below the autoresonant
threshold ($\epsilon < \epsilon_{\rm th}$), virtually no electrons
leave the well. In addition, the effect is still observed when the
laser frequency is mismatched with respect to the harmonic
oscillator frequency $\omega_0$. Even for a mismatch of $\pm
10\%$, the same number of electrons are ejected from the well.

As a second application, we consider a periodic array of quantum
wells, with a cosinusoidal confining potential $V_{\rm conf} =
-V_0 \cos(2\pi x/\lambda)$, where $\lambda$ is the width of each
well, and $V_0 = \mstar\omega_0^2 \lambda^2/2\pi$. Such periodic
superlattices can be practically realized as multilayer
semiconductor heterostructures \cite{Deveaud} and there have been
recent attempts at simulating these structures using Bose-Einstein
condensates trapped in optical lattices \cite{Sias}. Here, we
neglect the Hartree potential and consider noninteracting
electrons (this amounts to assuming $\eta \ll 1$). We take
$\gamma=0.002\omega_0$, $\alpha=-0.002\omega_0^2$, $\omega_0
t_0=300$, and $\epsilon = 0.2 \hbar\omega_0/L_{\rm ho}$.
Initially, each quantum well is occupied by a single electron,
represented by a minimum uncertainty packet. We use a periodic
computational box with spatial period equal to $3\lambda$.

\begin{figure}[htb]
\includegraphics[height=5.5cm, width=7.5cm]{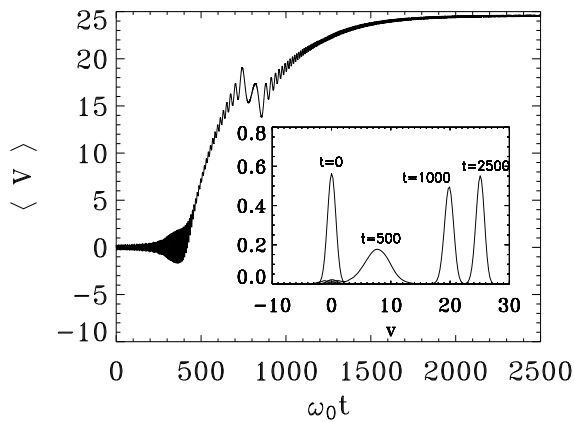}
\caption{\label{fig:3}Evolution of the average velocity for an
array of quantum wells with dc bias. Inset: Velocity distribution
of the electron population at $\omega_0 t=0$, 500, 1000, and 2500.
Velocity is measured in units of $\sqrt{\hbar\omega_0/\mstar}$.}
\end{figure}

The idea is to create an electron current by applying a suitable
laser pulse. From the previous study, we have learnt that the
electrons can be efficiently extracted from the well. By applying
a small constant (dc) electric field, we can force the electrons
to always leave the well from one side and thus create an electric
current. For this, we use a small dc field ($E_0 =
0.1\hbar\omega_0/eL_{\rm ho} \simeq 0.08$mV/nm) to polarize the
array of quantum wells \cite{Lee}. In Fig. 3, we plot the
evolution of the average velocity of the electron distribution
$\langle v \rangle = \int fv dv dx/\int f dv dx$.

\begin{figure}[htb]
\includegraphics[height=5cm, width=7.3cm]{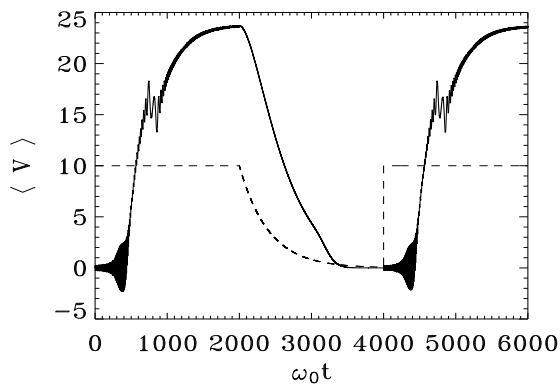}
\caption{\label{fig:4} Evolution of the average velocity for an
array of quantum wells with dc bias, in the case of two laser
pulses. The dashed line represents the evolution of the dc field,
in arbitrary units.}
\end{figure}

At $t=t_0$, a positive current starts building up and, by the end
of the run, the average velocity reaches a value $\langle v
\rangle \simeq 25$ in our normalized units. This value corresponds
to the asymptotic velocity reached by an electron subjected to a
constant electric field $E_0=0.1$ in the presence of a friction
coefficient $2\gamma=0.004$, i.e. $v_{\rm max} = E_0/2\gamma = 25$
(normalized units). This shows that all electrons have been
detrapped from the potential well and accelerated by the dc field.
The electron velocity distribution (Fig. 3, inset) confirms this
scenario.

In order to make the whole process reversible, one would need to
re-trap the electrons inside the potential well. This can be
achieved in the following way (Fig. 4): (i) after the current has
been generated, the oscillating pulse is switched off suddenly at
$\omega_0t = 2000$; (ii) at the same time, the dc electric field
$E_0$ is switched off adiabatically between $\omega_0t = 2000$ and
$\omega_0t = 4000$ (exponential decrease with time constant $\tau
=400\omega_0^{-1}$): during this phase, all electrons are
re-trapped and the current goes back to zero; (iii) at $\omega_0t
= 4000$, the dc electric field is switched on again and another
autoresonant laser pulse (identical to the first one) is used to
excite the current once again. This procedure can be repeated
several times, so that the electric current can be switched on and
off, and can even change sign if one changes the sign of the dc
field.

In summary, we presented numerical evidence that the electron gas
in wide nonparabolic quantum wells can be efficiently excited with
an autoresonant laser pulse. Nonlinear effects were triggered
using a relatively weak pulse, even when the laser frequency is
poorly tuned. These techniques could be used to achieve better
control of the electron dynamics in quantum solid-state devices.


\end{document}